\documentstyle[aps,prl,epsfig,array,floats]{revtex}
\begin{document}

\draft

\twocolumn[

\hsize\textwidth\columnwidth\hsize\csname@twocolumnfalse\endcsname

\title{Glassy behavior in systems with Kac-type step-function
interaction}
\author{Kok-Kiong Loh, Kyozi Kawasaki\footnotemark[2], Alan R. Bishop, Turab Lookman,
 Avadh Saxena}
\address{Theoretical Division, Los Alamos National Laboratory,
Los Alamos, NM 87545}
\author{J\"{o}rg Schmalian}
\address{Department of Physics and Astronomy and Ames Laboratory, Iowa
State University, Ames, Iowa 50011}
\author{Zohar Nussinov}
\address{Instituut-Lorentz for Theoretical Physics, Leiden University,
P.O.B. 9506, 2300 RA Leiden, The Netherlands}
\date{\today} \maketitle
\begin{abstract}
We study a system with a weak, long-range repulsive Kac-type
step-function interaction within the framework of a replicated
effective $\varphi^4$-theory. The occurrence of extensive
configurational entropy, or an exponentially large number of
metastable minima in the free energy (characteristic of a glassy
state), is demonstrated. The underlying mechanism of mesoscopic
patterning and defect organizations is discussed.
\end{abstract}
]

\footnotetext[2]{Permanent address: 4-37-9 Takamidai, Higashi-ku,
Fukuoka 811-0215, Japan.}

Competing interactions on different length scales cause in many
cases the emergence of an intermediate length scale where new
structures and inhomogeneities are formed. Examples are stripe
formation in doped Mott insulators\cite{stripe}, bubbles of
electronic states of high Landau levels in quantum Hall
systems\cite{QHE}, domains in magnetic multilayer
compounds\cite{magmtlyr}, and mesoscopic structures formed in
self-assembly systems\cite{selfassbl}. These systems typically
exhibit a multi-time-scale dynamics similar to the relaxation
found in glasses. The glassy behavior and the diverging relaxation
time are believed to be the result of the competition between the
interactions with different characteristic length
scales\cite{Nussinov99,Schmalian}; for example the macroscopic
phase separation is frustrated by the competing long-range
interactions\cite{Nussinov99,Kivelson}. Glassiness then arises
spontaneously in the absence of extrinsic disorder due to
self-generated randomness.

While there are various different scenarios for glassy behavior
such as the kinetic constraint where the diverging relaxation time
is purely of dynamic origin and occurs in a system with trivial
equilibrium properties, the central theme of this work is based
upon the random first order transition\cite{KTW89}, where
glassiness is attributed to an
 exponentially large number of metastable states, originally emphasized by
Kauzmann \cite{Kauzman}. The fact that configurational entropy is
needed for slow motions in glasses was first described by Gibbs
and DiMarzio \cite{GibbsDiMarzio}.
 Below a  crossover temperature $T_A$, an energy landscape-dominated,
``viscous'' long-time relaxation sets in due to an exponentially
large number of  metastable states, ${\cal N}$,  i.e., the
configurational entropy, $S_c=\ln{\cal N}$, becomes extensive.
This cross-over temperature $T_A$ is often associated with the
mode-coupling temperature at which the relaxation time or the
viscosity exhibits a power law divergence $|T-T_A|^{-\gamma}$,
within mode-coupling theory \cite{MCT}. Activation processes,
which are neglected in  mode-coupling theories, soften the sharp
transition into a crossover where  for $T<T_A$  free energy
barriers and thus transition rates  between the metastable states
remain finite. The configurational entropy decreases with
decreasing temperature, and becomes negative for $T<T_K$.  A
continuous (random first order) transition, the ``ideal'' glass
transition, occurs at $T_K$ to avoid $S_c<0$. The Kauzmann
temperature $T_K$ is the temperature at which the genuine
thermodynamic glass transition is expected, whereas the
experimentally observed glass transition occurs at $T_g>T_K$ which
depends on the cooling rate.

In Ref.~\cite{Schmalian}, it was  shown quite generally within the
framework of a replicated $\varphi^4$-theory that the competition
between short-range forces (favoring phase separation) and
long-range Coulomb interaction leads to an exponentially large
number of metastable states and self-generated glassiness. The
large phase space of fluctuations, which can lead to a fluctuation
induced first order transition\cite{Brazovskii}, was shown to,
alternatively, drive the system into an amorphous state, the {\em
stripe glass}. It is of particular interest to explore to what
extent {\em explicit} competition or frustration is necessary to
cause a glassy state and what kind of interactions support such a
state. A particularly interesting potential is the  Kac-type step
potential
\begin{eqnarray}
V(x)&=&\alpha^2 \gamma^D\phi(\gamma x),\label{Kac}
\end{eqnarray}
with $\phi(y)=1$ for $y\leq 1$ and zero otherwise.
Here, $\gamma$ controls the amplitude and range of the potential,
whereas the inverse length $\alpha$ characterizes the integral
strength $\int d^D x V(x) \propto \alpha^2$.
  In the van der Waals limit, $\gamma\rightarrow0$
   after the thermodynamic limit,  it is known that a
system of particles interacting through a potential given by
Eq.~(\ref{Kac}) can be described exactly by a mean-field theory
\cite{Grewe}, where $\alpha$ is related to the long-range force of
the van der Waals theory.  For the particular case of a step
potential, mean-field theory predicts a spinodal.
 This model system is of particular interest since it has been
used to study the glass formation and crystallization processes
\cite{Klein}.  Using Monte Carlo simulations, the appearance of
many metastable amorphous ``clump'' configurations was
demonstrated in Ref.~\cite{Klein}. However, it is difficult, if
not impossible, to enumerate the metastable states and to
determine the dependence of the number of local minima,  as a
function of the system size, in a Monte Carlo study. We find that
an analysis of a replicated $\varphi^4$-theory along the lines of
Ref.\cite{Schmalian} is useful as an alternative strategy.

In this paper we demonstrate that the Kac-type step function
interaction, Eq.~(\ref{Kac}), indeed causes the emergence of an
exponentially large number of metastable states and a {\em
self-generated glassy state}. In the limit of small but finite
$\gamma$ no actual frustration between different interactions is
necessary, which is different from the model studied in
Ref.\cite{Nussinov99,Schmalian}. We  demonstrate that glassiness
is due to multiple configurations of self-arrested defects, and
occurs, similar to Ref.\cite{Schmalian}, once the correlation
length of the system is slightly larger than the length scale of
mesoscopically modulated structures. The configurational entropy
behaves for $D=3$ like $S_c/V\propto \gamma^3$, where, $V$ is the
volume of the system. A model of glassy behavior with the Kac-type
long-range interaction, called a van der Waals glass, was
introduced and extensively developed, including a study of its
dynamics, in Ref.~\cite{kk02}, where it is argued that proximity
to the mean-field spinodal provides a long correlation length in
addition to the Kac potential range, and thus leads to frustration
and nonzero configurational entropy. This will be verified here
using the replica approach.

We start from the model Hamiltonian
\begin{eqnarray} {\cal H} &=& \frac{1}{2}\int
d^3x\left\{\left[\nabla\varphi(x)\right]^2+
r_0\varphi^2(x)+\frac{u}{2}\varphi^4(x) \right\} \nonumber\\
&&+ \int d^3x\int d^3x^\prime\varphi(x)\varphi(x^\prime)
V(x-x^\prime),\label{phi4}
\end{eqnarray}
with  $V(x)$ of Eq.~(\ref{Kac}). The usual equilibrium free energy
${\cal F}=-T\ln {\cal Z}$ is an outcome of an unconditional
average over the entire configuration space. It does not permit
the detection of local minima of the free energy in the
configuration space.  In Ref.~\cite{Monasson}, a
 replica approach is  proposed to overcome this limitation and
allows us to probe  the number of metastable states in models with
the  additional long-range interaction $V(x)$. We adopt the
self-consistent screening approximation to the replicated
$\varphi^4$-theory developed in Ref.~\cite{Schmalian}, where the
free energy ${\cal F}(m)$ of the replicated Hamiltonian is given
in terms of the regular correlation function $G(q)$ and the
correlation function
$F(q)\equiv\langle\varphi^a(q)\varphi^b(-q)\rangle$ between the
fields in different replicas. Here, $a \neq b$ are  the replica
indices and $m$ is the number of replicas. The quantity  $F(q)$
equals the Edwards-Anderson order-parameter,
$\lim_{t\rightarrow\infty}\langle\varphi(-q,t)\varphi(q,0)\rangle$,
which becomes nonzero in the glassy state, or when ergodicity is
broken.

 Within  the self-consistent screening
approximation, the free energy ${\cal F}(m)$ can be expressed as
\begin{eqnarray}
{\cal F}(m)=-\frac{T}{m}(\mbox{tr}\ln {\cal G}^{-1}+\mbox{tr}\ln
{\cal D}^{-1}),\label{fe}
\end{eqnarray}
and the configurational entropy $S_c$ is related to ${\cal F}(m)$
by
$S_c=\left.\frac{1}{T}\frac{d {\cal F}(m)}{dm}\right|_{m=1}$.
${\cal G}$ is the generalized correlation function matrix, defined
by
${\cal G}\equiv(G-F){\bf I}+F{\bf E},$
with the constant matrices ${\bf I}_{ab}=\delta_{ab}$ and ${\bf
E}_{ab}=1$. The symbol tr in Eq.~(\ref{fe}) includes the trace of
the replica and the momentum integral.  The momentum dependence of
the matrices is implied here but omitted for convenience. The
matrix ${\cal D}$ is related to ${\cal G}$ via the relation
${\cal D}^{-1}=(uT)^{-1}+\Pi$,
where $\Pi$ is the generalized polarization matrix defined as
$\Pi = (G\otimes G-F\otimes F){\bf I} + (F\otimes F){\bf E}$,
where $\otimes$ denotes convolution in Fourier space.  The
replicated  Schwinger-Dyson equation can be written as
${\cal G}^{-1}=G_0^{-1}{\bf I}+\Sigma, $
where $\Sigma$ is the self-energy and
\begin{eqnarray}
G_0(q)=\frac{1}{q^2+r+\tilde{V}(q)}, \label{GF}
\end{eqnarray}
with the renormalized mass $r = r_0+\int \frac{d^3q}{(2\pi)^3}\,
G$.
$\tilde{V}(q)$ denotes the Fourier transform of the potential
$V(x)$. Within  the self-consistent screening approximation the
self-energy has diagonal elements $\Sigma_G = 2G\otimes D_G$ and
off-diagonal elements $\Sigma_F =2F\otimes D_F$ in replica space,
where $D_G$ and $D_F$ being, respectively the diagonal and
off-diagonal elements of the matrix ${\cal D}$. These equations
form a closed set of self-consistent equations which enable us to
solve for $G$ and $F$, and then determine the configurational
entropy via
\begin{eqnarray}
S_{\rm c}=\int dq
\left\{S\left[\frac{F}{G}\right]-S\left[\frac{F\otimes F
}{(uT)^{-1}+G\otimes G}\right]\right\}. \label{Sconf}
\end{eqnarray}
Here, $S[x]\equiv-x-\ln(1-x)$.  It is obvious from the above
equation that $F\neq0$ implies $S_c\neq0$.

In the limit of small and large $\gamma$ it is possible to make
analytic progress. For large $\gamma$, i.e. short-range potentials
the gradient term in Eq.~(\ref{GF}) dominates and ${\tilde V}(q)$
can be neglected. No glassy state with finite Edwards-Anderson
parameter results. The situation is more interesting in the limit
of small $\gamma$, i.e. for long-range interactions. Using
$\tilde{V}(q)=\alpha^2 \psi(q/\gamma)$ with $\psi(z)=4
\pi(\sin(z)-z \cos(z))/z^3$, it follows  for $\gamma \ll \alpha$
that the gradient term in Eq.~(\ref{GF}) can  be neglected
compared to the Kac-type interaction. The long wavelength behavior
is dominated by the long-range interaction. The short-range
interaction becomes effectively local and has no characteristic
length scale anymore. The correlations are dominated by wave
vectors which minimize $\tilde{V}(q)$. Since $\psi(z)$ is minimal
(and negative) for  $z=z_0=5.76$, the dominant peak in the
correlation function occurs  at $q_0= z_0 \gamma$,
 independent of $\alpha$:
\begin{eqnarray}
G_0(q)\approx\frac{Z}{\xi^{-2}+\left(|{\bf q}|-q_0\right)^2}.
\label{GFapprox}
\end{eqnarray}
Here $Z=\frac{2}{c} (\gamma/\alpha)^2$ is the weight of the peak
with width characterized by $\xi^{-2}=Z(r+c \alpha^2)$, where
$c=\psi''(z_0)\approx 0.361$. This expansion around $q=q_0$
elucidates the correspondence with the analysis performed in
Ref.~\cite{Schmalian}. A  calculation along the lines of
Ref.~\cite{Schmalian} determines $T_{\rm A}$ as the temperature
where the  ratio of the correlation length $\xi$ and the
modulation length $l_0=2\pi/q_0$
 becomes larger than $2$. Inserting  Eq.~(\ref{GFapprox}) into
Eq.~(\ref{Sconf})
 yields for the configurational entropy
$S_{\rm c}=V C \gamma^3$,
 with $C\approx 6.81 \times 10^{-3}$. The longer range the
interaction, the smaller is the number of metastable states per
unit volume. As expected, glassiness disappears in the van der
Waals limit $\gamma \to 0$\cite{KKcmt}.

\begin{figure}
\centerline{\epsfig{file=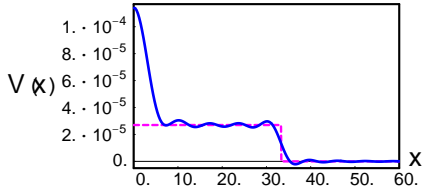, width=2.2in}}
\caption{Potential used in the numerical calculations (solid line)
and the step-function potential with $\gamma = 0.03$ and $\alpha=
1$ (dashed line).}
\label{pot}
\end{figure}

\begin{figure}
\centerline{\epsfig{file=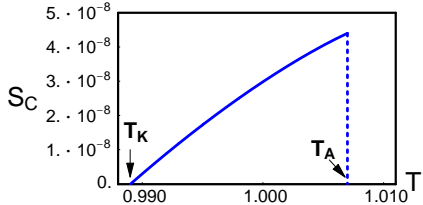, width=2.2in}}
\caption{Configurational entropy $S_c$ (within a replica theory)
as a function of temperature for $r_0=0.1494$ and $u=1.79$.}
\label{ce}
\end{figure}

\begin{figure}
\centerline{\epsfig{file=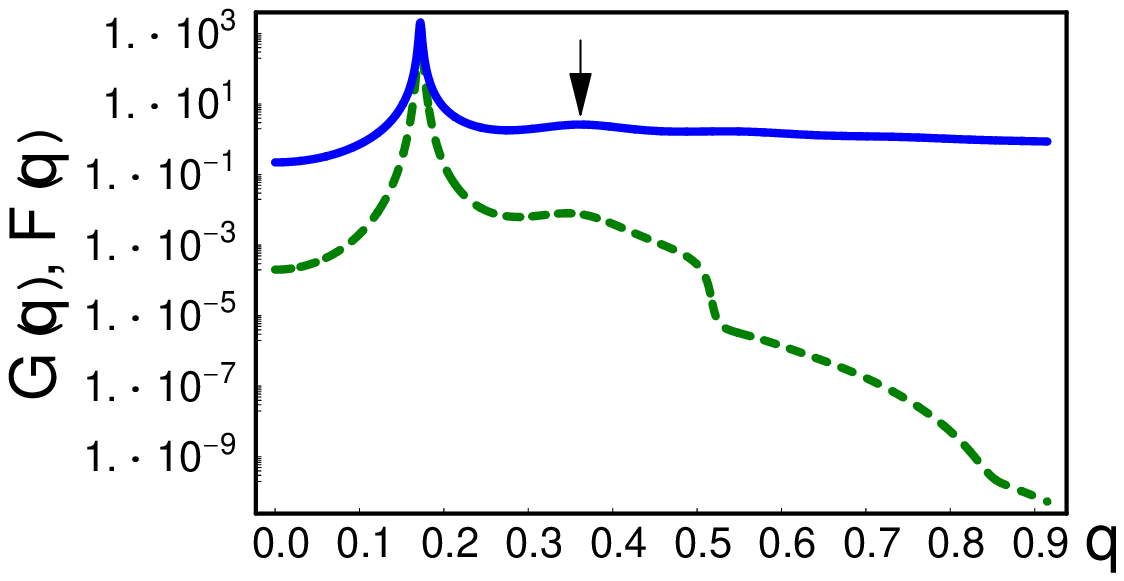, width=2.2in}}
\caption{Instantaneous correlation function  $G(q)$ (solid line)
and long-time correlation function $F(q)$ (dashed line) for $T=1$,
$r_0=0.1494$, and $u=1.79$. The arrow points at the position of
the first secondary peak.} \label{GFf}
\end{figure}

The set of self-consistent equations is solved numerically,
without assuming Eq.~(\ref{GFapprox}). A slightly modified
Kac-potential (see Fig.~\ref{pot}) has been used for convenience.
The major conclusions of this paper seem not affected by this
modification. In Fig.~\ref{ce}, we show the temperature dependence
of $S_c$ for $r_0=0.1494>0$ and $u=1.79$ (we use units where the
upper cutoff of the momentum integration is unity). The behavior
of $S_c$ matches identically the entropy crisis scenario of the
random first order transition theory~\cite{KTW89}. The
mode-coupling temperature $T_A$ and the Kauzmann temperature $T_K$
can be unambiguously identified. Furthermore,  we show in
Fig.~\ref{GFf} the correlation functions $G(q)$ and $F(q)$.  In
addition to the pronounced peak at $q_0$ we can also see higher
order structures in the instantaneous correlation function $G(q)$,
which become strongly suppressed  in the long-time correlation
function $F(q)$. The typical length scale which determines the
suppression of long-time correlations for larger $q$ is the defect
wandering length $\lambda$ which results from the off-diagonal
self-energy in replica space $\Sigma_{\rm
F}(q_0)=-\left(\frac{2}{\lambda}\right)^2$~\cite{Schmalian}.

Our analysis allows us to determine what kind of interactions can
cause a glassy state.  For small $\gamma$ the short-range term
will be effectively local.  Competing interactions of different
length scales can satisfy the glass-forming condition. Their
presence is, however, not necessary. We have just demonstrated
that nonzero configurational entropy can arise even if there is no
explicit competition between the interaction length scales. In
this case, not only does $\tilde{V}(q)$ possess a minimum at
$q_0$, but also $\tilde{V}(q_0)$ is negative. The system exhibits
a spinodal when $r_0>0$, i.e. for a purely repulsive interaction,
in distinction to the model discussed in
Ref.~\cite{Nussinov99,Schmalian}.  The glass-forming criterion
$\xi\geq2l_0$ can be fulfilled even in the absence of the gradient
term. We recall  that proximity to a mean-field spinodal provides
a large $\xi$ in addition to the Kac potential range, and thus
self-consistently leads to frustration. One can in fact ``map''
the van der Waals glass \cite{kk02} into the stripe
glass\cite{Schmalian}. It is thus possible to predict if
glassiness occurs in a system interacting through a mesoscale
potential $V(x)$ . For example, in the case of a Kac-type Gaussian
potential \cite{Stillinger} with $\phi(x)=\exp(-x^2)$, the Fourier
transform of this potential is monotonic. A modulation length
$l_0$ at the mesoscale does not occur and no glassy behavior is
anticipated in the absence of the short-range gradient term. This
agrees with the conclusion reached in Ref.~\cite{Klein}.  Both the
gradient term in the Hamiltonian and $r_0<0$ are necessary in
order to find similar glassiness at mesoscales.

From these considerations it seems to follow that the free energy
landscape of metastable states which causes glassiness originates
from configurations of mesoscopic defects. The origin of the
glassiness has been discussed in detail in Ref.~\cite{Schmalian}
in terms of a novel length scale, the {\em defect-wandering
length} $\lambda$, which depends on $\xi$ and $\l_0$.  In the
parameter region where the system acquires modulation at $l_0$,
the mesoscopically ordered state constitutes the global minimum of
the free energy ${\cal F}$, or the ground state.  With respect to
this ordered array of mesoscopic structures, excitations such as
'dislocations' are termed defects. The defect-wandering length
$\lambda$ is the distance on which a defect can move freely in the
lattice of the mesoscopic structure. When $\lambda<2l_0/3$
\cite{Schmalian}, the defects are pinned by the underlying lattice
of the mesoscopic structure. A distribution of such pinned defects
becomes a local minimum of ${\cal F}$. The organizations of
defects should be responsible for the exponentially large number
of metastable states ${\cal N}=\exp S_c$, where $S_c$ is
extensive. This picture is consistent with the findings of
Ref.~\cite{Klein} that the FCC arrangement of the clumps has lower
free energy than those of all frozen amorphous clump
configurations. In this case, the FCC clump configuration is an
ordered array of mesoscopic spherical structures.  The amorphous
clump phases are the organizations of defects about the ordered
lattice of these mesoscale spherical structures.

Another interesting question is whether glassiness is also
possible with a microscopic $l_0$ on the order of the hard core
radius of the atoms. The defect-pinning picture for the metastable
configurations described above is intuitively clear and consistent
with the 3D models where $l_0$ is mesoscopic. Nothing in the
argument forbids its application to a circumstance where $l_0$ is
the microscopic lattice constant. To study this in more detail, we
have investigated a 1D $\varphi^4$-model on a lattice in which the
frozen configurations of kinks are expected to give a large number
of metastable states. This occurs in the parameter regime where
the continuous variable $\varphi$ can be mapped to the 1D Ising
model with nearest-neighbor interaction, which has been shown to
have $S_c=0$ \cite{loh}.  A crude estimate of the number of
low-lying metastable states reflects {\em algebraic}, instead of
exponential size-dependence, leading to $\lim_{V \to \infty}
S_c/V=0$. While the effect of dimensionality is not clear, it is
thought not to change the system size dependence of the number of
low-lying metastable states from algebraic to exponential. We thus
conjecture that while pinning generates metastable states, the
internal organizations of the defects with respect to a
mesoscopically ordered state give rise to an exponentially large
number of these metastable states. An extensive $S_c$ does not
occur in the case of pinned defects about a microscopic lattice
because there are no internal degrees of freedom.

In summary, we have demonstrated glassiness (in the sense of a
finite configurational entropy) in a system interacting through a
Kac-type repulsive step-potential. We show that although the
mesoscopic modulation length $l_0$ is due to the interaction
potential alone, in contrast to the model proposed in
Ref.~\cite{Schmalian}, the underlying principle for glass
formation is identical, i.e. $\xi\geq 2l_0$.  In the absence of a
short-range interaction (the gradient term in the Hamiltonian)
this criterion for glassiness can only be fulfilled when
$\tilde{V}(q)$ possesses at least one minimum.  The concept of a
defect-wandering length is useful for understanding the source of
the exponentially large number of metastable states.  The picture
of the origin of the configurational entropy as the organizations
of defects in ordered arrays of mesoscopic structures is
consistent with the numerical findings in Ref.~\cite{Klein}.  The
glassiness we have found is due to configurations of mesoscopic
defects. Based on the analysis of a 1D Ising model, it is
speculated that only pinning of defects about the mesoscopic
pattern can give rise to exponentially large numbers of metastable
states.

KL acknowledges a Director's Fellowship at Los Alamos National
Laboratory. This research was  supported by the Department of
Energy, under contract W-7405-ENG-36 (KK) and W-7405-ENG-82 (JS).
An additional  partial support to KK by the Cooperative Research
under the Japan-U.S. Cooperative Science Program sponsored by
Japan Society of Promotion of Science is also gratefully
acknowledged.  JS acknowledges helpful discussions with P. G.
Wolynes and the hospitality of the Max Planck Institute for
Physics of Complex Systems, Dresden. We thank W. Klein for
discussions. This work was supported by the US Department of
Energy.


\begin{references}
\bibitem{KKcmt}If we scale up the
system size as $\gamma^{-D}$, glassiness still remains even in
this limit, the case effectively considered in Ref.~\cite{kk02}.
\end{references}
\end{document}